\documentstyle{article}

\def \cp89{{\elevenit CP Violation,} edited by C. Jarlskog (World Scientific,
Singapore, 1989)}
\def \hb87{{\elevenit Proceeding of the 1987 International Symposium on Lepton
and
Photon Interactions at High Energies,} Hamburg, 1987, ed. by W. Bartel
and R. R\"uckl (Nucl. Phys. B, Proc. Suppl., vol. 3) (North-Holland,
Amsterdam, 1988)}

\def \ichep72{{\elevenit Proceedings of the XVI International Conference on
High
Energy Physics}, Chicago and Batavia, Illinois, Sept. 6 -- 13, 1972,
edited by J. D. Jackson, A. Roberts, and R. Donaldson (Fermilab, Batavia,
IL, 1972)}

\def \ite{{\it et al.}}
\def \lkl87{{\it Selected Topics in Electroweak Interactions} (Proceedings of
the Second Lake Louise Institute on New Frontiers in Particle Physics, 15 --
21 February, 1987), edited by J. M. Cameron \ite~(World Scientific, Singapore,
1987)}
\def \ky85{{\it Proceedings of the International Symposium on Lepton and
Photon Interactions at High Energy,} Kyoto, Aug.~19-24, 1985, edited by M.
Konuma and K. Takahashi (Kyoto Univ., Kyoto, 1985)}

\def \prd#1#2#3{Phys. Rev. D {\elevenbf#1}, #2 (#3)}
\def \prl#1#2#3{Phys. Rev. Lett. {\elevenbf#1}, #2 (#3)}

\def \ptp#1#2#3{Prog. Theor. Phys. {\elevenbf#1}, #2 (#3)}

\def \si90{25th International Conference on High Energy Physics, Singapore,
Aug. 2-8, 1990}
\def \slc87{{\it Proceedings of the Salt Lake City Meeting} (Division of
Particles and Fields, American Physical Society, Salt Lake City, Utah, 1987),
ed. by C. DeTar and J. S. Ball (World Scientific, Singapore, 1987)}
\def \slac89{{\it Proceedings of the XIVth International Symposium on
Lepton and Photon Interactions,} Stanford, California, 1989, edited by M.
Riordan (World Scientific, Singapore, 1990)}
\def \smass82{{\it Proceedings of the 1982 DPF Summer Study on Elementary
Particle Physics and Future Facilities}, Snowmass, Colorado, edited by R.
Donaldson, R. Gustafson, and F. Paige (World Scientific, Singapore, 1982)}

\font\tenbf=cmbx10
\font\tenrm=cmr10
\font\tenit=cmti10
\font\elevenbf=cmbx10 scaled\magstep 1
\font\elevenrm=cmr10 scaled\magstep 1
\font\elevenit=cmti10 scaled\magstep 1

\font\ninerm=cmr9

\begin{document}
\renewcommand{\thefootnote}{\dag}
\begin{center}{\tenbf MODEL OF THE QUARK MIXING
MATRIX\footnote{\ninerm\baselineskip=11pt Presented by Mihir P. Worah at
DPF 92 Meeting, Fermilab, November, 1992.}\\}
\vspace{-1in}
\rightline{EFI 92-62}
\rightline{November 1992}
\bigskip
\vglue 2.3cm
{\tenrm JONATHAN L. ROSNER and MIHIR P. WORAH\\}
{\tenit Enrico Fermi Institute and Department of Physics, University
of Chicago\\}
\baselineskip=12pt
{\tenit Chicago, Illinois 60637,USA\\}
\vglue 0.8cm
{\tenrm ABSTRACT}
\end{center}
\vglue 0.3cm
{\rightskip=3pc
 \leftskip=3pc
 \tenrm\baselineskip=12pt
 \noindent
The structure of the Cabibbo-Kobayashi-Maskawa (CKM) matrix is analyzed from
the standpoint of a composite model. A model is constructed with three families
of quarks, by taking tensor products of sufficient numbers of spin-1/2
representations and imagining the dominant terms in the mass matrix to
arise from spin-spin interactions.  Generic results then obtained include the
familiar relation $|V_{us}| = (m_d/m_s)^{1/2} - (m_u/m_c)^{1/2}$, and a less
frequently seen relation $|V_{cb}| = \sqrt{2} [(m_s/m_b) - (m_c/m_t)]$.  The
magnitudes of $V_{ub}$ and $V_{td}$ come out naturally to be of the right
order.  The phase in the CKM matrix can be put in by hand, but its origin
remains obscure.
\vglue 0.6cm}
{\elevenbf\noindent 1. Introduction}
\vglue 0.1cm
\baselineskip=14pt
\elevenrm
The pattern of charge-changing weak transitions among quarks undoubtedly is a
reflection of deeper physics.  In the present article$^1$ we examine the form
this pattern might be expected to take if the underlying physics is that of a
composite system.  We construct a three-level quantum mechanical model of
composite quarks,  thereby obtaining an illustration of how the quark mixing
matrix, or Cabibbo-Kobayashi-Maskawa (CKM) matrix$^{2}$, might arise from some
underlying substructure.

A crucial problem$^3$ associated with the construction of realistic models is
to understand why, if the compositeness mass scale is very high, quarks and
leptons are so much lighter than this scale.  We assume that the dynamics is
such as to solve this problem. We shall imagine that the families of quarks and
leptons are the low-mass states (perhaps massless in some limit) of a
quantum-mechanical system whose other states lie at the compositeness scale.
We shall assume that this scale is independent of the physics giving rise to
any of the quark masses.

Because of our familiarity with quark models, we shall use a language which is
closely related to the nonrelativistic quark model.  We recognize that a proper
treatment of deeply bound composite systems may require a relativistic
treatment;  however, we find nonrelativistic discussions a convenient means of
counting states and of dealing with dynamically induced masses of the unwanted
(i.e., higher-lying) excitations.

\vglue 0.4cm
{\elevenbf\noindent 2. Model for three families of quarks}
\vglue 0.4cm
It so happens that one can form precisely three spin-1/2 levels with the
product of three spin-1/2 subunits and a spin-1 subunit.  The tensor product in
question is
$$
\frac{1}{2} \otimes \frac{1}{2} \otimes \frac{1}{2} \otimes 1 =
3 \left( \frac{1}{2} \right) \oplus 3 \left( \frac{3}{2} \right) \oplus
\frac{5}{2}~~~.
\eqno(1)
$$
We shall adopt such a model for quark and lepton families, regarding them as
the spin-1/2 members of the set (1).  We do not inquire here into the origin of
the subunits.  The spin-1 subunit may either arise as a result of the
composition of two spin-1/2 subunits in the manner mentioned above, or it may
be due to some internal orbital angular momentum.  In what follows we shall
label the spins of the three spin-1/2 subunits by ${\bf S}_i~(i = 1,2, 3)$ and
that of the spin-1 subunit by ${\bf L}$. We shall adopt the following basis for
description of our states. 1)  We imagine ${\bf S}_1$ and ${\bf S}_2$ to be
coupled to a total spin ${\bf S}_{12} \equiv {\bf S}_1 + {\bf S}_2$. 2)  Next,
we define ${\bf J}_{12} \equiv {\bf S}_{12} + {\bf L}$. 3)  Finally, we form
the total spin ${\bf J} = {\bf J}_{12} + {\bf S}_3$. We may then label our
three spin-1/2 states by $|S_{12},~J_{12} \rangle$.  The basis states are
$$
|0,~1 \rangle \equiv |u_0 \rangle~;~~~
|1,~1 \rangle \equiv |c_0 \rangle~;~~~
|1,~0 \rangle \equiv |t_0 \rangle~.
\eqno(2)
$$

We shall take our model Hamiltonian to consist of a constant term and linear
combinations of terms proportional to ${\bf S}_i \cdot {\bf S}_j$ and ${\bf L}
\cdot {\bf S}_i$.We examine the following simplified form of the mass matrix
for quarks,
$$
M(q) = M_0(J) + a {\bf S_1} \cdot {\bf S_3} + b {\bf S_2} \cdot {\bf S_3} +
c (1 + {\bf L} \cdot {\bf S_3})~~~,
\eqno(3)
$$
where $M_0(J) = {\rm const.} \times [J(J+1) - 3/4]$ is chosen so as to make all
states with $J > 1/2$ arbitrarily heavy, while $M_0(1/2) = 0$.

The mass operator (3) leads to the following mass matrix for spin-1/2 quarks:

$$
{\cal M}_{1/2} = \left[ \begin{array}{c c c}
0 & \alpha \sqrt{2} & \alpha \\
\alpha \sqrt{2} & \beta & \beta \sqrt{2} \\
\alpha & \beta \sqrt{2} & \gamma \\
\end{array} \right]~~~,
\eqno(4)
$$
where
$$
\alpha \equiv (a-b)/4~~~;~~~~\beta \equiv (2c - a - b)/4~~~;
{}~~~~\gamma = c
\eqno(5)
$$
With separate mass matrices (4) for up and down quarks, one has six parameters
with which to describe six quark masses. This theory has no CP violation.

Without a microscopic theory of CP violation, the best we can do in composite
models of the present variety is to put in phases by hand. We let $\alpha
\rightarrow -i\alpha ,\gamma \rightarrow -i\gamma$ in the mass matrix for the
up quarks. We then assume
$$
{\cal M}_{U} = \left[ \begin{array}{c c c}
0 & -i \alpha \sqrt{2} & -i \alpha \\
-i \alpha \sqrt{2} & \beta & \beta \sqrt{2} \\
-i \alpha & \beta \sqrt{2} & -i \gamma \\
\end{array} \right]~;~~~
{\cal M}_{D} = \left[ \begin{array}{c c c}
0 & \alpha' \sqrt{2} & \alpha' \\
\alpha' \sqrt{2} & \beta' & \beta' \sqrt{2} \\
\alpha' & \beta' \sqrt{2} & \gamma' \\
\end{array} \right]~,
\eqno(6)
$$
for $Q = 2/3$ and $Q = -1/3$ quarks, respectively. In the limit that $\alpha
\ll \beta \ll \gamma$ we obtain the following approximate CKM matrix elements:
$$
V_{us} \simeq i \left( \frac{-m_d}{m_s} \right)^{1/2}- \left( \frac{-m_u}{m_c}
\right)^{1/2}~~~;~~~~
V_{cb} \simeq \sqrt{2} \left( \frac{m_s}{m_b} -i \frac{m_c}{m_t} \right)~~~;
\eqno(7a,b)
$$
$$
V_{ub} \simeq i \frac{(-m_d m_s /2)^{1/2}}{m_b} - \frac{m_s}{m_b} \left(\frac{
-2 m_u}{m_c}\right)^{1/2} + \frac{(-m_u m_c /2)^{1/2}}{m_t} ~~~;
\eqno(7c)
$$
$$
V_{td} \simeq \frac{(-m_d m_s /2)^{1/2}}{m_b} +i \frac{m_c}{m_t} \left( \frac{
-2 m_d}{m_s} \right)^{1/2} + \frac{(-m_u m_c /2)^{1/2}}{m_t} ~~~.
\eqno(7d)
$$

We substitute a set of sample quark masses to see the implications of Eqs.~(7).
We choose$^4$
$
m_u = 5.2~{\rm MeV}~;~m_d = 9.2~{\rm MeV}~;~m_s = 194~{\rm MeV}~;~m_c =
1.41~{\rm GeV}~; ~m_b = 6.33~{\rm GeV}~;~m_t = 200~{\rm GeV}~.
$
These are values at a scale of 1 GeV. After rephasing to get to the
Wolfenstein parametrization$^5$ of the CKM matrix, we obtain
$
V_{us} = 0.226 ~;~V_{cb} = 0.044~;~
V_{ub} = 0.0051 - 0.0024~i~;~V_{td} = 0.0049 - 0.0024~i~;~
\rho = 0.51~;~\eta = 0.24~.
$

The above results are typical consequences of the expressions (7), and have
some interesting features.

1)  The magnitude of $V_{cb}$ is governed by the ratios of quark masses, rather
than their square roots.  As a result, one expects smaller typical values for
this quantity than for $V_{us}$.

2)  The orders of magnitude of $|V_{ub}|$ and $|V_{td}|$ are correct.

3)  The parameters in the mass matrices (6)  may be related to those in the
mass operator (3) via Eqs.~(5).  In the solution in which the signs of the
masses of the four heaviest quarks are all positive, $a,~b,$ and $c$ are all of
order $m_t$, with small splittings among these quantities responsible for the
masses of $u$ and $c$. Similarly, $a',~b'$, and $c'$ are all of order $m_b$,
with small splittings responsible for $m_d$ and $m_s$.  The presence of ${\bf
S}_3$ in nearly all the terms of the mass operator (3) then would suggest that
the spin-1/2 subunit carrying $S_3$ changes its identity under a
charge-changing weak transition.  However,  it seems difficult to preserve the
connection between ${\bf S}_3$ and the spin of the composite fermion.

4). The origin of the CP violating phase remains a mystery.
\vglue 0.5cm
{\elevenbf \noindent 3. Acknowledgments}
\vglue 0.3cm
This work was supported in part by the U. S. Department of Energy under Grant
No. DE FG02 90ER 40560.
\vglue 0.5cm
{\elevenbf\noindent 4. References \hfil}
\vglue 0.3cm
\begin{enumerate}

\item For more details, and complete references, see J. L. Rosner and M. Worah,
\prd{46}{1131}{1992}.

\item N. Cabibbo, \prl{10}{531}{1963};  M. Kobayashi and T. Maskawa,
\ptp{49}{652}{1973}.

\item G. 't Hooft, in {\elevenit Recent Developments in Gauge Theories}
(Carg\`ese Summer Institute, Aug. 26 - Sept. 8, 1979), ed. by G. 't Hooft
{\elevenit et al.} (Plenum, New York, 1980), p. 135.

\item H. Arason {\elevenit et al}., Phys. Rev. D (to be published)

\item L. Wolfenstein, \prl{51}{1945}{1983}.
\end{enumerate}
\end{document}